\documentclass[aps,prd,twocolumn,superscriptaddress,showpacs,amsmath,amssymb,amsthm,nofootinbib, preprintnumbers]{revtex4-2}
 \usepackage{amsmath}
\usepackage{graphicx}
\usepackage{epstopdf}
\usepackage{float}
\usepackage{hyperref}
\usepackage{color}
\usepackage[T1]{fontenc}
\usepackage[utf8]{inputenc}
\usepackage[toc,page]{appendix}
\usepackage[usenames,dvipsnames]{xcolor}
\usepackage[normalem]{ulem}
\usepackage{lipsum, babel}
\usepackage[justification=justified,format=plain]{caption}

\usepackage{subcaption}
\usepackage[utf8]{inputenc}
\usepackage{url}
\usepackage{xcolor}
\usepackage{physics}

\usepackage{hyperref}  

\usepackage{graphicx}% Include figure files
\usepackage{dcolumn}% Align table columns on decimal point
\usepackage{bm}% bold math
\newcommand{\be}{\begin{equation}}             %:skip:
\newcommand{\ee}{\end{equation}}               %:skip:
\newcommand{\ba}{\begin{eqnarray}}
\newcommand{\ea}{\end{eqnarray}}

\begin{document}

\title{New interpretation of the original charged BTZ black hole spacetime}

\author{Tom{\'a}{\v s} Hale}
\email{tomas.hale@utf.mff.cuni.cz}
\affiliation{Institute of Theoretical Physics, Faculty of Mathematics and Physics,
Charles University, V Holešovičkách 2, 180 00 Prague 8, Czech Republic}

\author{Brayden R. Hull }
\email[Corresponding Author]{b2hull@uwaterloo.ca}
\affiliation{Department of Applied Mathematics, University of Waterloo, Waterloo, Ontario, N2L 3G1, Canada}
\affiliation{Perimeter Institute for Theoretical Physics, Waterloo, ON N2L 2Y5, Canada}

\author{David Kubiz\v n\'ak}
\email{david.kubiznak@matfyz.cuni.cz}
\affiliation{Institute of Theoretical Physics, Faculty of Mathematics and Physics,
Charles University, V Holešovičkách 2, 180 00 Prague 8, Czech Republic}

\author{Robert B. Mann}
%\email{david.kubiznak@matfyz.cuni.cz}
\email{rbmann@uwaterloo.ca}
\affiliation{Department of Physics and Astronomy, University of Waterloo, Waterloo, Ontario, N2L 3G1, Canada}

\author{Jana Men{\v s}{\'i}kov{\'a}}
\email{jana.mensikova@matfyz.cuni.cz}
\affiliation{Institute of Theoretical Physics, Faculty of Mathematics and Physics,
Charles University, V Holešovičkách 2, 180 00 Prague 8, Czech Republic}

\date{\today}            % version 1.00; arxiv version 1

\begin{abstract}
In their seminal 1992 paper, Ba\~{n}ados, Teitelboim and Zanelli (BTZ) proposed a simple charged generalization of what is now known as the spinning BTZ black hole, the proposal being that a rotating metric can be supported by a `static vector' potential.
While with such an ansatz the Einstein equations are satisfied, and the corresponding energy-momentum tensor is divergence-less, the Maxwell equations do not (due to the special degenerate form of the corresponding field strength) hold. More recently, Deshpande and Lunin have proposed a generalized `Einstein--Maxwell' system which yields analytic rotating black holes in all odd dimensions. In this paper, we show that the original charged BTZ solution can be re-interpreted as a solution of the Deshpande--Lunin theory. Moreover, as we shall explicitly illustrate on an example of regularized conformal electrodynamics, similar construction also works for any non-linear electrodynamics in 3-dimensions. At the same time, all these spacetimes represent self-gravitating solutions of (NLE generalized) force-free electrodynamics.
\end{abstract}

\maketitle

\section{Introduction}
\label{sc:intro}

%According to the so-called \textit{no hair theorem}, black hole solutions of the Einstein-Maxwell equations can be fully characterized by their mass, electric charge, and angular momentum. 
 The quest for finding rotating and  electrically charged black hole solutions has a long history. 
While static (and charged) solutions of Einstein equations were discovered immediately after the invent of 
%publication of theory of 
general relativity, the rotating case was  much more difficult -- it was nearly 50 years after the introduction of general relativity that the famous Kerr solution \cite{Kerr:1963ud} was finally found in 1963. The corresponding charged and rotating solution was discovered soon after that, in 1965, by applying the (somewhat physically obscure) Newman--Janis transformation to a static charged black hole, thereby obtaining  the well-known rotating and charged Kerr--Newman solution in four dimensions \cite{Newman:1965tw}.

Interestingly, the task of extending such a solution to other spacetime dimensions turned out to be equally difficult. Vacuum rotating spacetimes in all higher dimensions were obtained by Myers and Perry in 1986 \cite{Myers:1986un}, and were later extended in 2004 to include 
$\Lambda$ \cite{Gibbons:2004js}, where
\be
\Lambda=-\frac{(d-1)(d-2)}{2\ell^2} 
\ee
is the cosmological constant. However, 
their charged versions, 
obeying the standard Einstein--Maxwell equations  derived from: 
%Lagrangian: 
\be\label{Maxwell}
{I}_{\mbox{\tiny EM}}=\frac{1}{16\pi}\int d^dx \sqrt{-g}(R-2\Lambda+4 {\cal L}_M)\,, 
\ee
where 
\be 
{\cal L}_M=-\frac{1}{2}{\cal S}\,,\quad {\cal S}=\frac{1}{2}F_{\mu\nu}F^{\mu\nu}\,,
\ee 
%\be\label{Maxwell}
%{I}_{\mbox{\tiny EM}}=\frac{1}{16\pi}\int \bigl[*(R-2\Lambda)-\frac{1}{2}F\wedge *F\bigr]\,, 
%\ee
and $F=dA$ is the Maxwell field strength in terms of the vector potential $A$, remain elusive to date. Whether or not this is due to the loss of conformal invariance of the Maxwell theory in $d\neq 4$ dimensions remains to be seen.

%However, a new interesting discovery was actually made in the realm of lower-dimensional gravity. 
Alternatively, in lower-dimensional gravity, 
 Ba\~{n}ados, Teitelboim, and Zanelli discovered in 1992 a new solution describing a rotating black hole in (2+1) dimensions with a negative cosmological constant \cite{Banados:1992wn, Banados:1992gq}, nicknamed the BTZ metric.
A charged version of this rotating solution was also proposed \cite{Banados:1992wn, Banados:1992gq};  surprisingly, it was characterized by a static vector potential,  
despite the spinning nature of the black hole. 
%Correct BTZ \cite{Clement:1995zt, PhysRevD.61.104013}.
However, although this rotating charged solution obeyed the Einstein equations, it was soon pointed out that it did not satisfy the vacuum Maxwell equations. A correct solution for the charged rotating BTZ black hole was subsequently found in 1995 \cite{Clement:1995zt} (see also \cite{Martinez:1999qi}) by applying a 
boost in the azimuthal direction
to the charged static solution, supplemented by a re-identification of the angle periodicity. For a review on the (2+1)-dimensional black hole physics see, e.g., \cite{Carlip:1995qv}.

Attempts at constructing analytic higher-dimensional charged rotating solutions continue to be of interest  \cite{Aliev:2006yk, Kunz:2006eh, Kunz:2007jq, Kolar:2015cha, Erbin:2016lzq, Andrade:2018rcx, Ortaggio:2023rzp}. In 2005, 
 a general charged and doubly spinning solution to $d=5$ minimal gauged supergravity
was found \cite{Chong:2005hr}, in which the action \eqref{Maxwell} is supplemented by  
a Chern--Simons term 
\be \label{CS}
{I}_{\mbox{\tiny CS}}^{\ (5)}=\frac{\lambda}{4\pi}\int A\wedge F\wedge F\,,
\ee 
whose coupling  constant must be fine tuned to $\lambda=\frac{1}{3\sqrt{3}}$.
The topological nature of this term yields the  ($\lambda$ independent) Einstein equations
\be\label{EE}
G_{\mu\nu}+\Lambda g_{\mu\nu}=8\pi T_{\mu\nu}\,, 
\ee
with the standard Maxwell energy-momentum tensor
\be\label{Tmunu}
T_{\mu\nu}=\frac{1}{4\pi}\Bigl(F_{\mu\alpha}F_{\nu}{}^\alpha-\frac{1}{4}g_{\mu\nu} F_{\alpha\beta}F^{\alpha\beta}\Bigr)\,, 
\ee
but the Maxwell equations pick up a current related to the self-interaction of the Maxwell field. Because of the former property, such a solution is `as close' to the Einstein--Maxwell theory `as possible'. 
Some attempts to go beyond the special coupling have been carried out \cite{Mir:2016dio, Blazquez-Salcedo:2016rkj}, but unfortunately, they do not seem to yield analytic solutions. At the same time, a similar trick, employing the `standard' topological Chern--Simons term:
\be\label{CSd}
I_{\mbox{\tiny CS}}^{(2n+1)}=\frac{\lambda}{4\pi}\int A\wedge \underbrace{F\wedge \dots \wedge F}_{n\mbox{\tiny -times}}\,, 
\ee
in $d=(2n+1)$ dimensions, does not yield (analytic) solutions  
either in $(2+1)$ dimensions \cite{Andrade:2005ur}  or in other  odd higher dimensions \cite{Deshpande:2024vbn} (see also \cite{Lu:2008ze, Allahverdizadeh:2010xx}).

Remarkably, an analytic generalization of the 
gauged supergravity solution \cite{Chong:2005hr}
in all odd higher dimensions (for arbitrary  couplings) was very recently found by Deshpande and Lunin \cite{Deshpande:2024vbn}. The key idea is to `generalize' the Chern--Simons term \eqref{CSd} to the action
\be \label{AHK}
I_{\mbox{\tiny DL}}=\frac{\lambda}{4\pi}\int A\wedge H \wedge K\,,
\ee 
in $d=(2n+1)$ dimensions, 
where $H=dB$, $K=dC$, with $B$ and $C$ are 
two new non-dynamical $(n-1)$-form fields,  and $\lambda$ is an arbitrary dimensionless coupling constant\footnote{ 
In the case of the standard Chern--Simons term \eqref{CSd} in $d=(2n+1)$ dimensions, 
the coupling constant $\lambda$ has dimensions $[\lambda]=L^{n-2}$, that is, it is dimensionless only in $d=5$ dimensions and in $d=3$ it has the units of inverse length. Contrary to this, in the Deshpande--Lunin theory, \eqref{AHK},  it is more natural to treat all the fields $A, B, C$ as well as the coupling constant $\lambda$ as dimensionless in any number of (odd) dimensions. 
}.

%\rbm{$R$ and $\Lambda$ both have units of $1/L^2$, so in $(2n+1)$ dimensions the gravity part of the action has
%units $L^{2n-1}$.  There is no coupling constant in the Maxwell part, so we must have $[A] =1$ in order for its field strength to be $1/L$ to get the same units as the gravity part. This is consistent with the action \eqref{CS}. Looking at \eqref{AHK},
%we see that the measure plus the 2 derivatives (one on $B$, the other on $C$) has units $L^{2n-1}$, which means that every other quantity is unitless. So $[\lambda]=1$. 
%}

The field equations obtained 
by adding \eqref{AHK} to the action \eqref{Maxwell}
  again yield the Einstein equations with the standard Maxwell energy momentum tensor \eqref{Tmunu}, together with the 
modified Maxwell equations:
\be\label{FJ} 
\nabla_\mu F^{\mu\nu}={\cal J}^\nu\,,
\ee 
where the current is given by 
\be \label{Jvia}
{\cal J}=\lambda\, *(H\wedge K)\,,
\ee 
in terms of the (auxiliary) field strengths $H$ and $K$. These quantities satisfy the two algebraic constraints
\be\label{algebraic} 
F\wedge H=0\,,\quad F\wedge K=0\,,
\ee 
in any odd dimension \cite{Deshpande:2024vbn}.  

It is the aim of this paper to show that the original charged BTZ solution  \cite{Banados:1992wn, Banados:1992gq} is a solution of the 
Deshpande--Lunin theory in $d=3$ dimensions.  We furthermore demonstrate  that any theory of Non-Linear Electrodynamics (NLE) yields exact solutions
to this theory.  Consequently, the original charged BTZ  metric is a viable metric that need not be discarded as an irrelevant or wrong solution.

%%%%%%%%%%%%%%%%%%%%%%%%%%
%%%%%%%%%%%%%%%%%%%%%%%%%%
\section{The Original Charged BTZ Solution}

\subsection{Basic properties}

The original proposal for a charged rotating black hole in $(2+1)$ dimensions was \cite{BTZ1992, RotatingBTZ1993} 
\be\label{BTZ1}
ds^2=-N^2f dt^2+\frac{dr^2}{f}+r^2(d\varphi+h dt)^2\,,
\ee
where the metric functions are given by 
\ba\label{BTZ0}
f&=&\frac{r^2}{\ell^2}+\frac{j^2}{r^2}-m-2Q^2\log(r/r_0)\,,\nonumber\\
h&=&\frac{j}{r^2}\,,\quad  N=1\,,
\ea
with
\be\label{BTZ2}
A= Q\log(r/r_0) dt
\ee
being the `static vector potential'.
The spacetime is characterized by three integration constants $m, Q$, and $j$, related to the mass, charge, and angular momentum, and an arbitrary length scale $r_0$. While it is possible to identify this scale with the cosmological radius $\ell$ (which consequently modifies the thermodynamic volume presented below), in what follows we shall treat $r_0$ as independent.  We   note that the length scale $r_0$ that appears in the potential and metric function need not be the same -- setting them equal   is a gauge choice.  More generally a second length scale $r_1$  could have been chosen in \eqref{BTZ2} instead of $r_0$.

The metric \eqref{BTZ1} satisfies the Einstein equations \eqref{EE},
%\be\label{EE}
%G_{\mu\nu}+\Lambda g_{\mu\nu}=8\pi T_{\mu\nu}\,, \quad \Lambda=-\frac{1}{\ell^2}\,,
%\ee
%that follow  from \eqref{Maxwell}, 
along with the necessary integrability condition
\be\label{Bianchi}
\nabla_\mu T^{\mu\nu}=0\,, 
\ee
which follows from the Bianchi identity.  However, as previously noted \cite{Clement:1995zt, Martinez:1999qi} (see also a recent discussion in \cite{Maeda:2023oei}), the Maxwell equations 
\be\label{vacuum ME}
\nabla_\mu F^{\mu\nu}=0\,, 
\ee
are not satisfied.  For this reason, the  spacetime \eqref{BTZ1}--\eqref{BTZ2} was  denoted as a wrong solution and disregarded, with  the `correct' charged BTZ black hole later  constructed by
a `boosting technique' \cite{Clement:1995zt, Martinez:1999qi}.
 Interestingly, although such a solution is `easy to write' in the Kerr-like coordinates, it becomes rather cumbersome in the original BTZ coordinates (see Appendix~\ref{app:a}).

Let us look at the `failure of the Maxwell equation' a bit more closely.  Using the definition of the electromagnetic energy-momentum tensor \eqref{Tmunu}, 
the above integrability condition \eqref{Bianchi} implies 
\be\label{force free}
0=4\pi \nabla_\mu T^{\mu\nu}=F^\nu{}_{\alpha}\nabla_\mu F^{\mu\alpha}\,, 
\ee
 where in the second equality we have used the `electromagnetic Bianchi identity', $dF=0$. 
Now, if $F_{\mu\nu}$ had an `inverse', the integrability condition \eqref{force free}  would then imply the vacuum Maxwell equations. However, in our case $F_{\mu\nu}$ is degenerate, and the Maxwell equations are only satisfied with non-trivial current on the r.h.s., namely: 
\be\label{ME}
\nabla_\mu F^{\mu\nu}={\cal J}^\nu\,, 
\ee
where 
\be\label{J}
{\cal J}=-\frac{2jQ}{r^4}\partial_\varphi\,. 
\ee
The above represents a current in the $\partial_{\varphi}$ direction penetrating the spacetime. It is this current that allows for a rotating solution with a `static' electromagnetic field. Note also that in the non-rotating case, $j\to0$, the current vanishes and we recover the standard charged static BTZ black hole. 

Remarkably,   equations \eqref{force free} and \eqref{ME}, together imply
 \be
 F_{\mu \nu}\mathcal{J}^{\nu}=0 \label{ffecond}
 \ee
which is a characteristic of {\em force-free electrodynamics}  \cite{Brennan:2013kea}, relevant for the Blandford--Znajek effect  of electromagnetic extraction of energy from a rotating black hole \cite{Blandford:1977ds}. Furthermore, as stated above, our electromagnetic tensor is degenerate which is another necessary condition of force-free electrodynamics and it is equivalent to $\mathcal{P}=0$ \cite{PhysRevE.56.2181}. 
The original charged BTZ spacetime thus represents a solution of `backreacting' force free electrodynamics.\footnote{Note that in our case, the force-free electrodynamics is electrically dominated since $\mathcal{S}=-\frac{Q^2}{r^2}=-E^2$. However, in plasma physics, the magnetically dominated (i.e. $\mathcal{S}>0$) case is relevant.}  
Moreover, as we shall discuss now, it is also a solution of the Deshpande--Lunin theory introduced above.
%\footnote{
%\tcr{{\bf If we want to include this observation, we should probably also include V Witzany as an author, I think.} Using the following identity, valid for any form $p$-form $\omega$ and any vector $X$:
%\be 
%X\cdot *\omega=*(\omega\wedge X)\,,
%\ee 
%one finds that 
%\be 
%{\cal J}\cdot F= (*d*F)\cdot F=*(F\wedge *d*F)=\pm *(F\wedge H\wedge K)=0\,,
%\ee 
%on behalf of the algebraic equations \eqref{}. Thence, any solution of Deshpande--Lunin theory is also a solution of force free electrodynamics. 
%}  
%}

\subsection{New interpretation of the original BTZ solution}

It is easy to verify that 
the original charged BTZ spacetime \eqref{BTZ1}--\eqref{BTZ2} 
solves the full set of Deshpande--Lunin  equations of motion \eqref{FJ}--\eqref{algebraic} in $(2+1)$ dimensions, provided we set\footnote{Interestingly, the solution for scalars $B$ and $C$ is not unique. For example, instead of the above \eqref{BC}, one may consider 
\be
B\to B/b\,,\quad C\to -r C \int \frac{b}{r^2}dr \nonumber
\ee
for arbitrary dimensionless function $b=b(r)$.  
 Note also that, since the Lagrangian in \eqref{AHK} is parity-odd, one of $B$ or $C$ must be in fact a pseudoscalar.
}
\be\label{BC}
B=\frac{t}{\lambda r} \,,\quad 
C=-\frac{2jQ}{r}\,. 
\ee
In particular, such fields automatically obey both algebraic equations \eqref{algebraic}, as well as give rise to the current \eqref{J} via \eqref{Jvia}. Namely, we have 
\be 
H=dB=\frac{dt}{\lambda r}-\frac{tdr}{\lambda r^2}\,,\quad K=dC=\frac{2jQ}{r^2}dr\,,
\ee 
and thence 
\ba 
{\cal J}&=&\lambda *(H\wedge K)=\frac{2jQ}{r^3}*(dt\wedge dr)\nonumber\\
&=&{-\frac{2 j^2 Q}{  r^4}dt-\frac{2 j Q}{r^2}d\varphi}\,,
\ea 
which, upon raising the index, yields the current \eqref{J}. 
  We have thus embedded the original charged BTZ solution in the Deshpande--Lunin theory. We note  the non-perturbative feature of this solution, encoded in the $1/\lambda$ behavior of the auxiliary fields.

\begin{center}
    \includegraphics[scale=0.42]{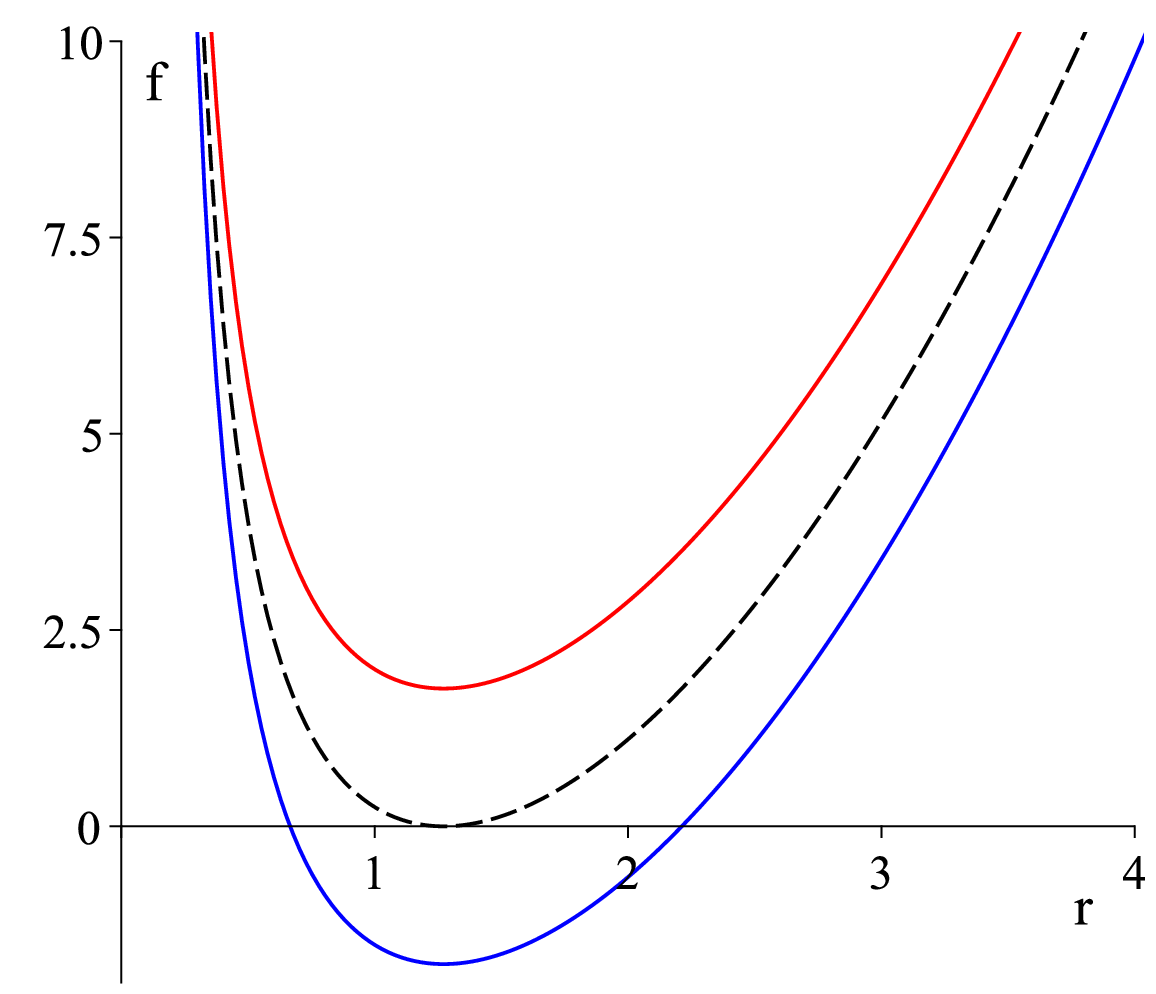}
    \captionof{figure}{{\bf Black hole horizons.} We display the metric function $f$ as a function of the mass parameter for fixed $\ell=1=Q=j=r_0$. The red curve corresponds to a charged point mass with 
    $m < m_E$,
    the dashed black curve to $m=m_E$ (an extremal black hole with single degenerate horizon), and the blue curve to  $m=2m_E$ (a non-extremal black hole with two horizons). }
    \label{fig1}
\end{center}

 Having `rehabilitated' the original charged BTZ solution, we can now proceed and study its basic properties. 
In particular, in a certain range of parameters, the solution will describe a black hole with up to two horizons. Namely, solving $f(r_E)=f'(r_E)=0$, we find that the radius $r_E$ of an extremal black hole is located at 
\be
r_E=\frac{\ell}{\sqrt{2}}\sqrt{Q^2+\sqrt{Q^4+4j^2/\ell^2}}\,, 
\ee  
and corresponds to the mass $m_E$:
\be
m_E\equiv \frac{r_E^2}{\ell^2}+\frac{j^2}{r_E^2}-2Q^2 \log(r_E/r_0)\,. 
\ee
For $m>m_E$ (with other parameters fixed), we then have a black hole with 2 horizons, while no horizons are present for $m<m_E$, as displayed in Fig.~\ref{fig1}. This latter case corresponds to a rotating charged point mass; see \cite{Akbar:2007zz} for a more detailed analysis.

The solution has a curvature singularity at $r=0$, where (in the presence of nontrivial charge $Q$) the Ricci scalar
\be
R=\frac{6}{\ell^2}-\frac{2Q^2}{r^2} 
\ee
diverges.

We can also construct the thermodynamic charges 
\ba 
M&=&\frac{m}{8}\,,\quad T=\frac{f'(r_+)}{4\pi}
=\frac{r_+^4-j^2\ell^2-Q^2\ell^2r_+^2}{2\pi r_+^3 \ell^2}\,,
\nonumber\\
S&=&\frac{\pi r_+}{2}\,,\quad \Omega=\frac{j}{r_+^2}\,,\quad 
J=\frac{j}{4}\,,
\nonumber
\\
\phi&=&-\frac{Q}{2}\log(r/r_0)\,,\\
V&=&\pi r_+^2\,,\quad P=\frac{1}{8\pi \ell^2}\,,\quad \Pi_{r_0}=\frac{Q^2}{4 r_0}\,,\nonumber
\ea 
and show that these obey the extended first law
\be
\delta M=T\delta S+\phi \delta Q+\Omega \delta J+ V\delta P+\Pi_{r_0}\delta r_0\,, 
\ee
together with the standard Smarr relation
\be
0=TS-2PV+\Omega J+\Pi_{r_0} r_0\,.
\ee 
 Remarkably, none of the above quantities depends on the coupling $\lambda$; 
 thermodynamic variables associated with
 the fields $B$ and $C$ do not seem to enter the thermodynamic laws.

\section{Exact solutions in non-linear electrodynamics}
\label{sc:NLE}

Let us now demonstrate that a similar construction also works 
 when the Maxwell Lagrangian is replaced with that of any NLE.
For concreteness, we focus on the case of recently constructed regularized conformal electrodynamics \cite{Kubiznak:2024ijq}; the general case follows straightforwardly.

\subsection{Deshpande--Lunin-NLE theory}

We generalize the Deshpande--Lunin
theory to NLE via the following action: 
\be
I_{\mbox{\tiny DL-NLE}}=\frac{1}{16\pi}\int d^dx\sqrt{-g}(R-2\Lambda+4{\cal L})+\frac{\lambda}{4\pi} \int A\wedge H\wedge K\,, 
\ee
where we have replaced the Maxwell Lagrangian, ${\cal L}={\cal L}_{\mbox{\tiny M}}=-\frac{1}{2}{\cal S}$, with an `arbitrary' function  
\be\label{LofS}
{\cal L}={\cal L}({\cal S})
\ee 
of the electromagnetic invariant ${\cal S}$. 
This yields  the Einstein equations \eqref{EE} with the modified electromagnetic energy-momentum tensor:
\be\label{TmunuNLE} 
T^{\mu\nu}=-\frac{1}{4\pi}\bigl(2F^\mu{}_\alpha F^{\nu\alpha}{\cal L}_{\cal S}-{\cal L}g^{\mu\nu}\bigr)\,,
\ee 
together with the NLE field equations:
\be\label{MaxwellNLE}
\nabla_\mu D^{\mu\nu}={\cal J}^\nu\,,\quad {\cal J}=\lambda * ( H \wedge K)\,,
\ee
where we have denoted 
\be
D_{\mu\nu}\equiv -2 {\cal L}_{\cal S}F_{\mu\nu}\,,\quad {\cal L}_{\cal S}\equiv \frac{\partial {\cal L}}{\partial {\cal S}}\,, 
\ee
along the Deshpande--Lunin algebraic equations 
\be\label{DHKNLE}
F\wedge H=0\,,\quad F\wedge K=0\,,
\ee
which remain unchanged.
 
We may also generalize the force-free electrodynamics to theories of NLE. Namely, we simply demand that 
\be 
\nabla_\mu T^{\mu\nu}=0\,,\quad dF=0\,,
\ee 
be satisfied in NLE. It is straightforward to show that these equations, together with the form of the electromagnetic stress tensor \eqref{TmunuNLE} and the (first) NLE field equation \eqref{MaxwellNLE}, imply 
\be 
0=  F^\nu{}_\alpha \nabla_\mu D^{\mu\alpha}=  F^\nu{}_\alpha {\cal J}^\alpha \,,
\ee 
with the latter meaning that the (NLE generalized) Lorenz force vanishes. As we shall see, the novel metrics constructed below satisfy both the Deshpande--Lunin-NLE equations of motion as well as the equations of NLE force-free eletrodynamics.

In particular, we shall focus on $d=3$ dimensions and consider the recently proposed {\em Regularized Conformal Electrodynamics} (RegConf) \cite{Kubiznak:2024ijq} for which the  Lagrangian reads: 
\ba\label{eq34}
{\cal L}_{\mbox{\tiny RC}}&=&-2\beta\alpha^3\Bigl(s+\frac{s^2}{2}+\ln(1-s)\Bigr)\,, \nonumber\\
s&\equiv&\Bigl(-\frac{{\cal S}}{\alpha^4}\Bigr)^{\frac{1}{4}}\in (0,1)\,.
\ea

The theory is characterized by two parameters $\alpha$ and $\beta$ whose dimensions are $[\alpha^2]=1/L=[\beta^2]$.  
The large-$\alpha$ limit  is equivalent to the small-$s$ limit of \eqref{eq34}. Namely, we get  
\ba
{\cal L}_{\mbox{\tiny RC}}&=&
-2\beta\alpha^3\Bigl(s+\frac{s^2}{2} - s - \frac{s^2}{2} - \frac{s^3}{3} + \cdots \Bigr)\nonumber\\
&=&   2\beta\alpha^3 \frac{s^3}{3}+\cdots  \nonumber \\
&\approx&  \frac{2}{3}\beta\alpha^3 \Bigl(-\frac{{\cal S}}{\alpha^4}\Bigr)^{\frac{3}{4}} 
=  \frac{2}{3}\beta \left( -{\cal S}\right)^{\frac{3}{4}}\,, 
\ea
indicating that   ($d=3$) conformal electrodynamics \cite{Martinez2007} 
\be \label{LC}
{\cal L}_{\mbox {\tiny C}}=\frac{2}{3}\beta (-{\cal S})^{3/4}\,,
\ee 
is recovered in the limit 
$\alpha\to \infty$.

As shown in \cite{Kubiznak:2024ijq}, RegConf theory is a natural continuation of the four-dimensional {\em Regularized Maxwell} (RegMax) theory \cite{Tahamtan:2020lvq, Hale:2023dpf} to three dimensions. It breaks the conformal symmetry of conformal electrodynamics \eqref{LC} by introducing a `minimal' regularization of the field strength, and is exceptional in that it admits simple analytic self-gravitating accelerating black hole solutions in three dimensions.

\subsection{Novel solution in Regularized Conformal Electrodynamics}
\label{IIIB}

 Similar to the Maxwell case, 
the solution can be written in the following form:  
\ba
f_{\mbox{\tiny RC}}&=&f_{\mbox{\tiny RC}}^{(0)}+\frac{j^2}{r^2}\,,\quad h=\frac{j}{r^2}\,,\quad N=1\,,\nonumber\\
A_{\mbox{\tiny RC}}&=&A_{\mbox{\tiny RC}}^{(0)}\,,
\ea
in terms of the static metric function $f_{\mbox{\tiny RC}}^{(0)}$ and static vector potential $A_{\mbox{\tiny RC}}^{(0)}$ \cite{Kubiznak:2024ijq}: 
\ba 
f_{\mbox{\tiny RC}}^{(0)}&=&
\frac{2\alpha Q^2}{\beta}-m-4Q\alpha^2 r+\frac{r^2}{\ell^2}\nonumber\\
&&+4\alpha^3\beta r^2 \log\Bigl(\frac{\alpha r+Q/\beta}{r\alpha}\Bigr)\,,\nonumber\\
A_{\mbox{\tiny RC}}^{(0)}&=&-\frac{\alpha Q^2}{\beta^2(\alpha r+Q/\beta)}dt\,,
\ea
with the  auxiliary fields still given by \eqref{BC}.

 The solution is characterized by the current \eqref{J} and the asymptotic electric charge 
\be
\frac{1}{2\pi}\int *D=Q  
\ee
and the thermodynamic quantities   
\ba 
M&=&\frac{m}{8}\,,\quad S=\frac{\pi r_+}{2}\,,\nonumber\\
T&=&\frac{f_{\mbox{\tiny RC}}'(r_+)}{4\pi}=\frac{r_+}{2 \pi  \ell^2}-\frac{\alpha ^2 Q (Q+2 \alpha  \beta  r_+)}{\pi  (Q+\alpha  \beta  r_+)}\nonumber\\
&&\quad -\frac{j^2}{2 \pi  r_+^3}+\frac{2 \alpha ^3 \beta  r_+}{\pi} \log \Bigl(\frac{Q+\alpha\beta r_+}{\alpha  \beta  r_+}\Bigr)\,,\nonumber\\
\Omega&=&\frac{j}{r_+^2}\,,\quad 
J=\frac{j}{4}\,,\quad \phi=\frac{\alpha  Q^2}{2 \left(\beta  Q+\alpha  \beta ^2 r_+\right)}\,,\nonumber\\
V&=&\pi r_+^2\,,\quad P=\frac{1}{8\pi \ell^2}\,,\\
\Pi_\alpha&=&\frac{3}{2}\beta\alpha^2r_+^2\log\Bigl(\frac{\alpha\beta r_++Q}{\alpha\beta r_+}\Bigr)\nonumber\\
&&\quad +\frac{Q(Q^2-3Q\alpha\beta r_+-6\alpha^2\beta^2r_+^2)}{4\beta(\alpha\beta r_++Q)}\,,\nonumber\\
\Pi_\beta&=&\frac{1}{2}\alpha^3r_+^2\log\Bigl(\frac{\alpha\beta r_++Q}{\alpha\beta r_+}\Bigr)\nonumber\\
&&\quad -\frac{Q\alpha(Q^2+Q\alpha\beta r_++2\alpha^2\beta^2r_+^2)}{4\beta^2(\alpha\beta r_++Q)}\,.\nonumber
\ea 
%\ba 
%M&=&\frac{m}{8}\,,\quad S=\frac{\pi r_+}{2}\,,\nonumber\\
%T&=&\frac{f_{\mbox{\tiny RC}}'(r_+)}{4\pi}\nonumber\\
%&=&-\frac{j^2}{2 \pi  r_+^3}+\frac{r_+}{2 \pi  \ell^2}+\frac{2 \alpha ^3 \beta  r_+}{\pi} \log \left(\frac{Q+\alpha\beta r_+}{\alpha  \beta  r_+}\right)\nonumber\\
%&&\quad -\frac{\alpha ^2 Q (Q+2 \alpha  \beta  r_+)}{\pi  (Q+\alpha  \beta  r_+)}\,,\nonumber\\
%\Omega&=&\frac{j}{r_+^2}\,,\quad 
%J=\frac{j}{4}\,,\nonumber\\
%\phi&=&\frac{\alpha  Q^2}{2 \left(\beta  Q+\alpha  \beta ^2 r_+\right)}\,,\nonumber\\
%V&=&\pi r_+^2\,,\quad P=\frac{1}{8\pi \ell^2}\,,\\
%\Pi_\alpha&=&\frac{3}{2}\beta\alpha^2r_+^2\log\Bigl(\frac{\alpha\beta r_++Q}{\alpha\beta r_+}\Bigr)\nonumber\\
%&&\quad +\frac{Q(Q^2-3Q\alpha\beta r_+-6\alpha^2\beta^2r_+^2)}{4\beta(\alpha\beta r_++Q)}\,,\nonumber\\
%\Pi_\beta&=&\frac{1}{2}\alpha^3r_+^2\log\Bigl(\frac{\alpha\beta r_++Q}{\alpha\beta r_+}\Bigr)\nonumber\\
%&&\quad -\frac{Q\alpha(Q^2+Q\alpha\beta r_++2\alpha^2\beta^2r_+^2)}{4\beta^2(\alpha\beta r_++Q)}\,.\nonumber
%\ea 
They obey the following extended first law and Smarr relations:
\ba
\delta M&=&T\delta S+\phi \delta Q+\Omega \delta J+V\delta P+\Pi_\alpha \delta \alpha+\Pi_\beta \delta \beta\,,\quad\\ 
0&=&TS+\Omega J-2PV-\frac{1}{2}\Pi_\alpha \alpha-\frac{1}{2}\Pi_\beta \beta\,.
\ea

\subsection{General solution for  any NLE}

More generally one can repeat the same construction for any NLE coupled to the 
Deshpande--Lunin theory. This solution is  characterized by the following functions:
\ba
f&=&f_0+\frac{j^2}{r^2}\,,\quad h=\frac{j}{r^2}\,,\quad N=1\,,\nonumber\\
A&=&A_0=-\phi_0dt\,, 
\ea
where $f_0=-m+\frac{r^2}{\ell^2}+\tilde f$ and $A_0$ are the corresponding static metric function and the static vector potential, respectively. The function $\tilde f$ depends on the particular choice of
NLE term
in \eqref{LofS}.

The Maxwell field strength is $D^{\mu\nu}=F^{\mu\nu}_{\mbox{\tiny M}}$, and the current \eqref{J} and
the Deshpande--Lunin fields 
\eqref{BC} remain unchanged. Without explicit knowledge of $\tilde f$ we obtain the thermodynamic quantities
\ba
M&=&\frac{m}{8}\,,\quad T=\frac{f'(r_+)}{4\pi}\,,\quad S=\frac{\pi r_+}{2}\,,\nonumber\\
\Omega&=&\frac{j}{r_+^2}\,,\quad 
J=\frac{j}{4}\,,\quad Q=\frac{1}{2\pi}\int *D\,,\nonumber\\
\phi&=&\frac{1}{2}\phi_0(r_+)\,,\quad 
V=\pi r_+^2\,,\quad P=\frac{1}{8\pi \ell^2}\,, 
\ea
together with additional potentials associated with (dimensionfull) couplings of a given NLE, obeying 
\be
\delta M=T\delta S+\phi \delta Q+ \Omega \delta J+V \delta P+\dots\,,
\ee   
where the dots represent additional variations of dimensionful parameters/couplings in the solution.

In particular, for $d=3$ conformal electrodynamics, \eqref{LC}, we have 
\ba
f_0&=& -m+\frac{4Q^3}{3r\beta^2}+\frac{r^2}{\ell^2}\,,\nonumber\\
A_0&=&-\frac{Q^2}{r\beta^2}dt\,,
\ea
generalizing the solution of \cite{Cataldo:2000we, Gurtug:2010dr, Cataldo:2020cxm}.
In addition to the above thermodynamic quantities, we also have a dimensionful coupling $\beta$ and the corresponding potential 
\be
\Pi_\beta=-\frac{Q^3}{3\beta^2 r_+}\,. 
\ee
These quantities then satisfy 
\ba
\delta M&=&T\delta S+\phi \delta Q+ \Omega \delta J+V \delta P+\Pi_\beta \delta \beta\,,\nonumber\\
0&=&TS+\Omega J-2PV-\frac{1}{2}\Pi_\beta \beta\,.
\ea  
It is straightforward to check that this solution, together with its thermodynamics, can be recovered upon taking the limit $\alpha\to \infty$, applied to the  solution in section~\ref{IIIB}.
.

\section{Summary}

We have rehabilitated the original charged BTZ spacetime  \cite{Banados:1992wn, Banados:1992gq}, known to be   peculiar in that it features a rotating black hole geometry supported with a static vector potential.
Although the metric satisfies the Einstein equations with the standard electromagnetic energy--momentum tensor, it only obeys the Maxwell equations with a non-trivial current on the right hand side. 
The corresponding electromagnetic field obeys the equations of force-free electrodynamics  sourced by a current derived from the
Deshpande--Lunin term \eqref{AHK}. The spacetime thus can be embedded in the Deshpande--Lunin theory and presents a viable rotating and charged black hole solution (which in addition is significantly simpler than the `correct' charged and rotating BTZ black hole spacetime).  
 The same construction also works 
for any  non-linear electrodynamic theory coupled to gravity provided the action
\eqref{AHK} is retained.

 It remains to be seen whether such a construction can also be extended to accelerated black holes of \cite{Kubiznak:2024ijq}, or even more interestingly to black holes in four (and even higher) dimensions.

%\tcg{In this paper we make the following observation: although the static vector potential \eqref{BTZ2} presented as a solution for the rotating charged BTZ black hole in \cite{RotatingBTZ1993} does not solve the dynamical Maxwell equations \eqref{vacuum ME}, the current \eqref{ME} that appears on the r.h.s. can be sourced by a Chern-Simons term in the action, that fits into the theory proposed in \cite{Deshpande:2024vbn}. It can thus be reinterpreted as a true solution of all the equations of motion in this new context. We note that this solution then becomes solution of \eqref{force free}, which we identify as the equation of force-free electrodynamics with the non-trivial current \eqref{ME}.\\
%In \ref{sc:NLE} we show how this construction can be extended to any theory of non-linear electrodynamics to include sources. These are generated by the Chern- Simons term, which imposes the satisfaction of two algebraic equations for the two auxiliary scalar fields.\\
%For a comparison of the discussed solution \eqref{BTZ2} with the `correct' solution in the original formulation derived in \cite{Clement:1993kc}, we obtain the form of the latter in the original BTZ coordinates in \ref{app:a}.
%}

\appendix

\section{Charged and spinning BTZ black hole}
\label{app:a}

The correct (`current-less') charged and spinning BTZ black hole was obtained in 
\cite{Clement:1995zt, Martinez:1999qi}, 
 by performing a `boost transformation' on 
the non-rotating charged seed. Such a solution is canonically written 
in  boosted (Kerr-like) coordinates.
Here we rewrite it in the `standard' BTZ coordinates, for a comparison to the solution in the main text. 

\subsection{Boosting technique}

We begin by reviewing the `boosting technique', which can be appplied to any charged static solutions, including those in NLE.
 Consider a static charged solution of the following form:
\ba
ds^2&=&-Fdt ^2+\frac{d\rho^2}{F}+\rho^2d\varphi^2\,,\nonumber\\
A&=&\phi dt\,.\quad 
\ea
Here, we decompose
\be
F=\frac{\rho^2}{\ell^2}-M-E(\rho)\,,
\ee
where $E(\rho)$ represents the `electromagnetic sector contribution' to the metric function, and  
$\phi=\phi(\rho)$ is the corresponding electrostatic potential, with $\rho$ playing the role of the radial coordinate.

Applying the following boost
\cite{Lemos:1994xp, Lemos:1995cm, Hennigar:2020drx}: %LEMOS199546,PhysRevD.54.3840}:
\begin{equation} \label{eq:ein_boosts}
    t \rightarrow \Xi t-a \varphi, \quad \varphi \rightarrow \frac{a t}{\ell^2}-\Xi \varphi, \quad \Xi^2=1+\frac{a^2}{\ell^2}\,,
\end{equation}
we obtain the solution 
\ba
ds^2&=&-F(\Xi dt-a d \varphi)^2+\frac{\rho^2}{\ell^4}(a dt-\Xi \ell^2 d \varphi)^2+\frac{d\rho^2}{F}\,, \nonumber\\
A&=&\phi(\Xi dt-a d\varphi)\,,
\ea
written in the {Kerr-like coordinates 
typically used in the literature. We  rewrite this in the BTZ coordinates
\ba
ds^2&=&-\frac{F\rho^2}{\Xi^2\rho^2-a^2 F}dt^2+\frac{d\rho^2}{F}\nonumber\\
&&+(\Xi^2\rho^2-a^2F)\Bigl(d\varphi+\frac{\Xi a(F\ell^2-\rho^2)}{\ell^2(\Xi^2\rho^2-a^2F)}dt\Bigr)^2 \quad  
\ea
by completing the square with respect to the angular coordinate. 

Assuming that $E(\rho)$ falls off at infinity faster than $\rho^2/\ell^2$, the periodicity of the new angular coordinate $\varphi$ can be identified with $2\pi$. This can be seen by employing the coordinate transformation
\be\label{transf}
 r^2\equiv \Xi^2\rho^2-a^2F(\rho)\,,
\ee
upon which the metric becomes
\be
ds^2=-\frac{F\rho^2}{r^2}dt^2+\frac{dr^2}{Fr'^2}+r^2\Bigl(d\varphi+\frac{\Xi(\rho^2-r^2)}{ar^2}dt\Bigr)^2,\quad\  
\ee
where $r'=\frac{dr}{d\rho}$. Comparing this to the standard  form \eqref{BTZ1}, namely:
}
\be
ds^2=-N^2fdt^2+\frac{dr^2}{f}+r^2(d \varphi+hdt)^2\,,
\ee
we must identify
\be\label{Nhf} 
f=Fr'^2\,,\quad N=\frac{\rho}{rr'}\,,\quad h=\frac{\Xi(\rho^2-r^2)}{ar^2}\,.
\ee

\subsection{Vacuum}
The simplest case to examine explicitly is the vacuum case, for which $E(\rho)=0$, that is, 
\be
F=\frac{\rho^2}{\ell^2}-M  
\ee 
implying that the transformation 
\eqref{transf} now reads
\be
r^2=\rho^2+a^2M\quad \Rightarrow \quad \rho=\sqrt{r^2-a^2M}\,. 
\ee
This in turn gives the following functions
\ba
N&=&1\,,\quad h=-\frac{a\Xi M}{r^2}\,,\nonumber\\
f&=&\frac{r^2}{\ell^2}-\frac{(2a^2+\ell^2)M}{\ell^2}+\frac{M^2a^2\Xi^2}{r^2}  
\ea
and so upon identifying 
\be
j=-a\Xi M\,,\quad m=\frac{(2a^2+\ell^2)M}{\ell^2}\,, 
\ee
we recover the standard uncharged spinning BTZ black hole \eqref{BTZ1}--\eqref{BTZ0}. 

%\ba 
%& N^{\phi} = & -\frac{\Xi a M }{R^2} \nonumber \\
%& N = & 1 \nonumber \\
%& F = & \frac{R^2}{l^2}-\frac{\left(2 a^2+l^2\right) M}{l^2}+\frac{M^2 a^2\left(a^2+l^2\right)}{l^2 R^2}
%\ea 
%which can be re-written as
%\ba 
%& N^{\phi} = & -\frac{J }{2 R^2} \nonumber \\
%& N = & 1 \nonumber \\
%& F = & \frac{R^2}{l^2}-\Tilde{m}+\frac{J^2}{4  R^2}
%\ea 
%for 
%\[2J = \Xi a M \]
%\[\frac{\left(2 a^2+l^2\right) M}{l^2} = \Tilde{m}. \]
%Considering $ \Omega \equiv g_{t \phi}/g_{\phi \phi}$
%we have 
%\be 
%\Omega=-\frac{a \Xi M}{R^2}
%\ee 
%and there as $R \to \infty$$\Omega \to 0$, where $R \to R_+$$ \Omega \to J $. 

\subsection{Maxwell}
In the Maxwell case we have $E(\rho) = -2 Q^2 \log \left(\rho / \rho_0\right)$. The radial transformation then reads  
 
\begin{equation}
r^2=a^2 M+\rho^2+2 a^2 Q^2 \log(\rho/\rho_0)\,. \
\end{equation}
Due to the logarithm present in this equation the inversion for $\rho$ is more difficult now, but nonetheless it has a solution of the form:
\begin{equation}
    \rho^2 = Q^2 a^2 \mbox{W}(x)\,,\quad x=\frac{{\rho_0}^2}{Q^2a^2}\exp\Bigl( \frac{r^2-a^2 M}{Q^2 a^2}\Bigr)
\end{equation}
where $\mbox{W}(x)$ is the Lambert-W function.
%\tcr{\bf Are you sure Brayden -- Maple finds something else for me!} \textcolor{cyan}{\bf Jana: Mathematica gave me the same thing as Brayden got.} \tcb{Brayden: After simplification you obtain what I posted here, Maple just doesn't simplify immediately.}
In turn, this yields the charged and rotating BTZ solution in the standard form via \eqref{Nhf}.
Obviously, the resultant solution contains Lambert functions and is quite complex and visually unappealing. In particular, we recover
\be
N=\frac{\rho^2}{\rho^2+a^2Q^2}\,, 
\ee 
which no longer equals unity.   This is to be compared to the `simple solution' \eqref{BTZ1}--\eqref{BTZ2} studied in the main text.

\section*{Acknowledgements}

 This work was supported in part by the Natural Sciences and Engineering Research Council of Canada. Research at the Perimeter Institute is supported by the Government of Canada through the Department of Innovation, Science and Economic Development and by the Province of Ontario through the Ministry of Colleges and Universities.
D.K. is grateful for support from GAČR
23-07457S grant of the Czech Science Foundation and the Charles University Research Center Grant No. UNCE24/SCI/016. We would like to thank Vojt{\v e}ch Witzany and David Kofro{\v n} for useful remarks on our paper.  D.K. would also like to thank Miok Park and the Institute for Basic Science (IBS) in Daejeon for their hospitality, where part of this work was completed.

\bibliographystyle{JHEP}
\bibliography{cited_only}

\end{document}